\documentclass{Interspeech2024}
\usepackage{authblk} 
\usepackage{amsmath}
\usepackage{array} 
\usepackage{cite} 
\usepackage{multirow}
\usepackage{graphicx}
\usepackage[ruled,linesnumbered]{algorithm2e}
\usepackage{authblk}

\interspeechcameraready

\title{Disentangling Age and Identity with 
a Mutual Information Minimization Approach for Cross-Age Speaker Verification }

\name[affiliation={*1}]{Fengrun}{Zhang}
\name[affiliation={*2}]{Wangjin}{Zhou}
\name[affiliation={1}]{Yiming}{Liu}
\name[affiliation={\#1}]{Wang}{Geng}
\name[affiliation={1}]{Yahui}{Shan}
\name[affiliation={1}]{Chen}{Zhang}
\address{
  $^1$Kuaishou Technology, Beijing, China\\
  $^2$Graduate School of Informatics, Kyoto University, Sakyo-ku,Kyoto, Japan 
  }
\email{zhangfengrun@kuaishou.com, zhou@sap.ist.i.kyoto-u.ac.jp, gengwang@kuaishou.com}
\begin{document}\maketitle
\renewcommand{\thefootnote}{\fnsymbol{footnote}}
\footnotetext{* Equally contributed.}
\footnotetext{\# Corresponding author.}

\begin{abstract}
There has been an increasing research interest in cross-age speaker verification~(CASV). However, existing speaker verification systems perform poorly in CASV due to the great individual differences in voice caused by aging. 
In this paper, we propose a disentangled representation learning framework for CASV based on mutual information~(MI) minimization. In our method, a backbone model is trained to disentangle the identity- and age-related embeddings from speaker information, and an MI estimator is trained to minimize the correlation between age- and identity-related embeddings via MI minimization, resulting in age-invariant speaker embeddings. 
Furthermore, by using the age gaps between positive and negative samples, we propose an aging-aware MI minimization loss function that allows the backbone model to focus more on the vocal changes with large age gaps. 
Experimental results show that the proposed method outperforms other methods on multiple Cross-Age test sets of Vox-CA.

\end{abstract}

\section{Introduction}
With the remarkable advancement of deep learning, current X-Vector based models~\cite{snyder2018x,desplanques20_interspeech,chen23o_interspeech} and margin-based loss functions ~\cite{deng2019arcface,zhao2022multi} have achieved excellent performance in automatic speaker verification~(ASV). Despite the impressive success of general ASV, most systems are not robust and are easily affected by various complex factors in reality, including speaker-to-microphone distance, language and speaking style. While increasing research attention has been attracted to the above factors~\cite{qin20_interspeech,thienpondt2022tackling,gonzalez2019limits}, aging factors on ASV are rarely studied due to the insufficiency of relevant data in the past.

Cross-Age Speaker Verification~(CASV) aims to verify the same speaker's recordings across different ages.
In recent years, \cite{tawara2021age, hechmi2021voxceleb} provide partial age labels of speakers from VoxCeleb~\cite{nagrani17_interspeech,chung18b_interspeech} by manual labeling. \cite{qin22_interspeech} adopt the face age estimation method to automatically conduct age labels for Voxceleb, providing possibilities for CASV task. 
\begin{figure}[t]
  \centering
  \includegraphics[width=\linewidth]{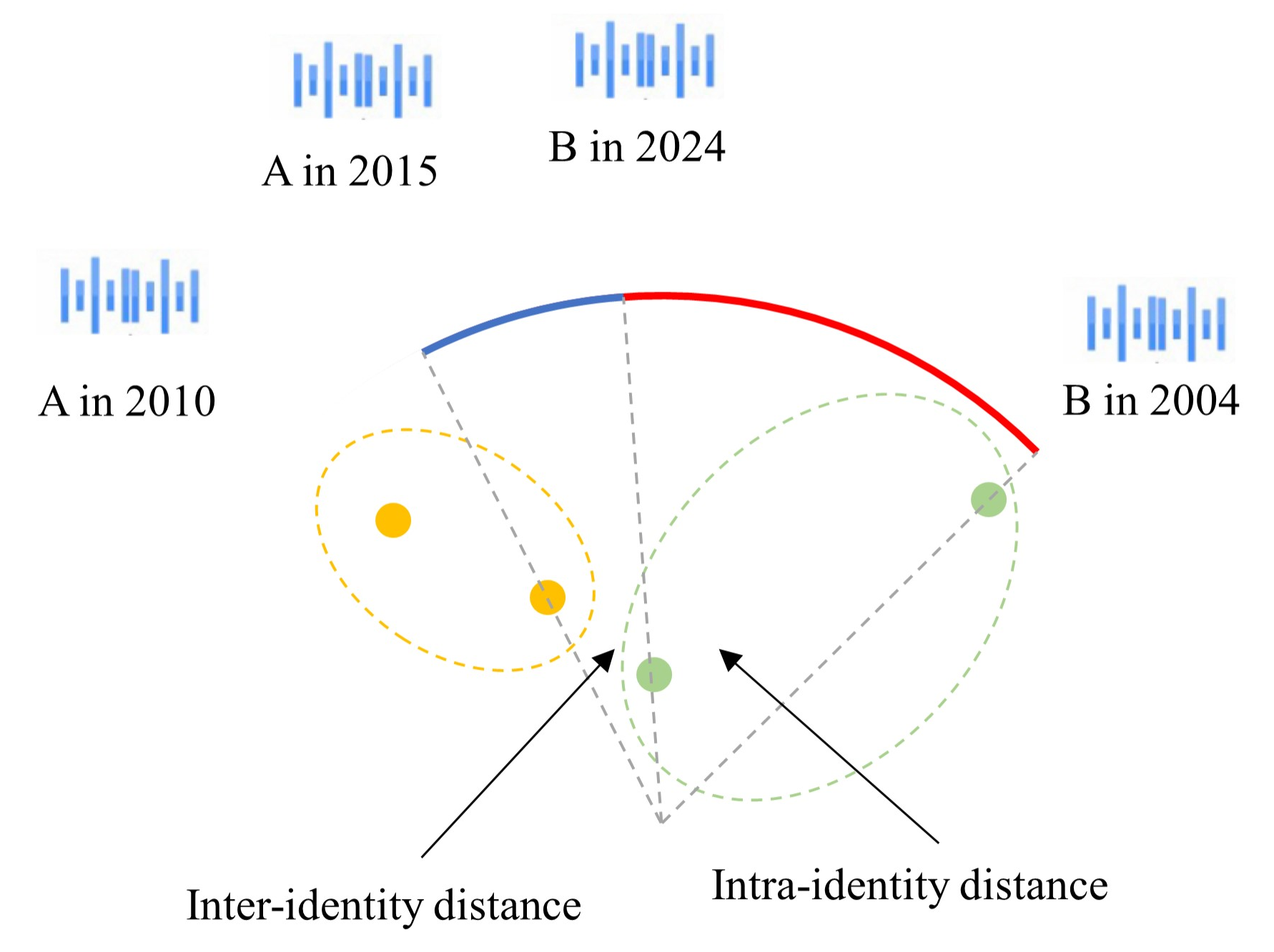}
  \caption{A typical example for CASV.
  }
  \label{fig1}
  \vspace{-0.4cm}
\end{figure}
The cross-age scenes are challenging because of the increasing intra-identity distance for the same speaker's voice caused by aging. Physiologically, as humans age, their vocal organs~(i.e., vocal tract and vocal folds) change, resulting in variations in acoustic characteristics~(i.e., F0, formant and loudness)~\cite{singh23d_interspeech, mueller1997aging}. In a word, the aging factors result in significant intra-identity variations.
A typical example is illustrated in~Fig.\ref{fig1} that speaker embeddings have marked variations within the same identity across different ages while those of different identities are more similar, raising challenges in current ASV systems. While voice aging is a complex process affected by intrinsic and extrinsic factors, the changes of measurable acoustic parameters have similar trends~\cite{reubold2010vocal}. Hence, it is feasible and significant to disentangle the speaker representation into age embedding and age-invariant identity embedding for CASV.

Existing disentangling method in CASV directly eliminates age information from speaker representation via an age classification loss and gradient reversal~\cite{qin22_interspeech}. However, the gradient reversal tends to confuse age information rather than recognize and disentangle it from identity information. Moreover, while making age information indistinguishable, gradient reversal can only learn age information, rather than effectiveness on speaker representation of aging factors.

In this paper, we propose a mutual information~(MI) based approach to learn age-invariant speaker embeddings. Specifically, our method consists of a backbone model and an MI estimator. 
The relationship between the two modules is similar to the generative adversarial network~\cite{goodfellow2014generative}. The MI estimator is trained to measure the MI between age and identity embeddings from the backbone model while the backbone model is guided to minimize the MI between these two embeddings by the MI estimator.
The MI minimization~(MIM) results in identity-related embeddings containing as little age information as possible, approaching age-invariant speaker embeddings. Moreover, we propose a novel aging-aware~(AA) MI minimization loss function to let the backbone model pay more attention to the influence of aging factors on speaker representation, leading to better age-invariant speaker embeddings.
\begin{figure*}[t]
  \centering
  \includegraphics[width=\linewidth]{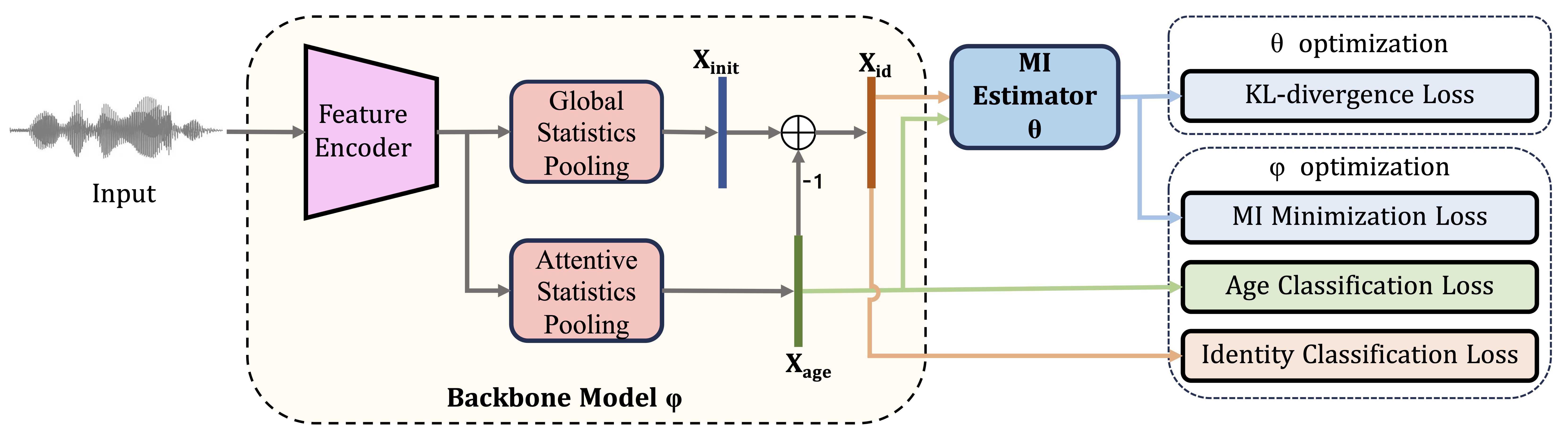}
  \caption{Proposed model structure.} 
  \label{fig2}
\end{figure*}

The major contributions are summarized as follows:
\begin{itemize}
    \item We propose a novel disentangling method based on minimizing the mutual information between age and age-invariant speaker embeddings.
    \item We present an aging-aware mutual information minimization loss function to better capture aging factors and learn age-invariant speaker embeddings.
    \item We demonstrate the effectiveness of the proposed method with experiments conducted on the Vox-CA dataset. Experimental results show relative improvement over the state-of-the-art~(SOTA) method over 1.53\% in EER and 4.79\% in minDCF, respectively. 
\end{itemize}

\section{Method}
As illustrated in Fig.~\ref{fig2}, our method consists of two modules. The backbone model extracts the initial embedding \( \boldsymbol{x_{\text{init}}} \) and disentangles it into age embedding \( \boldsymbol{x_{\text{age}}} \) and age-invariant identity embedding \( \boldsymbol{x_{\text{id}}} \). The mutual information (MI) estimator learns to estimate the MI and guides the backbone model to minimize the MI between \( \boldsymbol{x_{\text{age}}} \) and \( \boldsymbol{x_{\text{id}}} \) via back-propagation.

\subsection{Representation Disentanglement}
The backbone model is designed to extract age embedding $ \boldsymbol x_{age}$ and age-invariant identity embedding $ \boldsymbol x_{id}$. The feature encoder extracts feature maps $ \boldsymbol x'\in\mathbb{R}^{c\times f \times t}$ from  an input audio $ \boldsymbol x \in\mathbb{R}^{1\times T}$. As $ \boldsymbol x'$ represents high-level information, we adopt global statistics pooling to capture global speaker information $ \boldsymbol x_{init} \in\mathbb{R}^{d}$  and attentive statistics pooling $ \boldsymbol x_{age} \in\mathbb{R}^{d}$ to capture age-related information. These two pooling layers lead to initial disentanglement of age information.

Motivated by previous works~\cite{qin22_interspeech,hou2021disentangled}, we model $ \boldsymbol x_{age}$ and $ \boldsymbol x_{id}$ as statistically independent variables. 
Since the initial embedding $ \boldsymbol x_{init}$ represents speaker information, which consists of both age information and age-invariant information, we define the disentanglement as:
\begin{equation}
    \label{eq1}
 \boldsymbol x_{init}=\boldsymbol x_{age}+\boldsymbol x_{id}
\end{equation}

After disentanglement, we obtain the identity-related embedding $\boldsymbol{x}_{id}$ through a simple subtraction. We then use ArcFace loss $\mathcal{L}_{\mathrm{id}}$ for identity classification and softmax loss $\mathcal{L}_{\mathrm{age}}$ for age classification to learn the corresponding embeddings.

\begin{equation}
    \label{eq9}
\mathcal{L}_{id}=-\frac{1}{N}\sum_{i=1}^{N}\log\frac{e^{s\cdot\cos\left(\delta_{y_i}+m\right)}}{e^{s\cdot\cos\left(\delta_{y_i}+m\right)}+\sum_{j\neq y_i}e^{s\cdot\cos\delta_{j}}}
\end{equation}
\begin{equation}
    \label{eq10}
\mathcal{L}_{age}=-\frac{1}{N}\sum_{i=1}^{N}\log\frac{e^{z_i}}{\sum_{j=1}^Je^{z_j}}
\end{equation}
where \( y_i \) and \( z_i \) are the corresponding logits of the ground truth identity and age labels. Through this multi-task learning framework, we ensure that \( \boldsymbol x_{id} \) contains abundant identity-related information and \( \boldsymbol x_{age} \) contains abundant age-related information.

Different from identity classification, where different identities vary greatly, ages with few gaps are indistinguishable in real scenes. Therefore, the target of \( \mathcal{L}_{age} \) is to map \( \boldsymbol x_{age} \) into its corresponding age groups.


\subsection{Mutual Information Minimization}
\subsubsection{Basic Optimization}
MI estimation has been widely used in deep learning to constrain correlation between variables~\cite{wang2021learning,hou2021disentangled,li23r_interspeech}. We consider it also promising to disentangle age-related and age-invariant information from initial speaker embeddings.
The MI between $ \boldsymbol x_{age}$  and  $ \boldsymbol x_{id}$ is defined as :
\begin{equation}
    \label{eq2}
I(\boldsymbol x_{age};\boldsymbol x_{id})=\int_{\boldsymbol x_{age}\times \boldsymbol x_{id}}\log\frac{{p}({ \boldsymbol x_{age},  \boldsymbol x_{id}})}{{{p}({ \boldsymbol x_{age}}) {p}({ \boldsymbol x_{id}})}}d{p}({ \boldsymbol x_{age},  \boldsymbol x_{id}})
\end{equation}
where ${p}({ \boldsymbol x_{age},  \boldsymbol x_{id}})$ is the joint probability distribution while ${p}({ \boldsymbol x_{age}})$  and ${p}({ \boldsymbol x_{id}})$ are the marginals.

For MI minimization, the common practice is to estimate and minimize the upper bound of MI. In this paper, we adopt the sampled Contrastive Log-ratio Upper Bound (CLUB)~\cite{cheng2020club} as the basic MI minimizing function, which is better in the bias-variance estimation trade-off and more effective for MI minimization. As the conditional distribution $ {p}({ \boldsymbol x_{age}|{{ \boldsymbol x_{id}}}})$ is unknown, a variational distribution $ {q_\theta}({ \boldsymbol x_{age}|{{ \boldsymbol x_{id}}}})$ with parameter $ \theta$ modeled by a neural network is used to approximate real distribution $ {p}({ \boldsymbol x_{age}|{{ \boldsymbol x_{id}}}})$. The upper bound of MI is defined as:
\begin{equation}
    \label{eq3}
\begin{aligned}
\operatorname{I}(\boldsymbol x_{age}; \boldsymbol x_{id}):=& \mathbb{E}_{p({ \boldsymbol x_{id}},{ \boldsymbol x_{age}})}[\log q_{\theta}({ \boldsymbol x_{age}}|{ \boldsymbol x_{id}})]  \\
&-\mathbb{E}_{p({ \boldsymbol x_{id}})}\mathbb{E}_{p({ \boldsymbol x_{age}})}[\log q_{\theta}({ \boldsymbol x_{age}}|{ \boldsymbol x_{id}})]
\end{aligned}
\end{equation}
With sample pairs $\{(\boldsymbol{x}_{age},\boldsymbol{x}_{id})\}_{i=1}^{N}$, the MI minimization objective function for the backbone model is defined as:
\begin{equation}
    \label{eq4}
\mathcal{L}_{MIM}=\frac{1}{N}\sum_{i=1}^{N}\Big[\log q_{\theta}(\boldsymbol{x}_{age}^{i}|\boldsymbol{x}_{id}^{i})-\log q_{\theta}(\boldsymbol{x}_{age}^{k_{i}^{\prime}}|\boldsymbol{x}_{id}^{i})\Big]
\end{equation}
Where $(\boldsymbol{x}_{age}^{k_{i}^{\prime}}|\boldsymbol{x}_{id}^{i})$ is the negative pair sampled from the same training batch with the condition ${k_{i}^{\prime}} \neq i$.
Meanwhile, the MI estimator is trained to  minimize the KL-divergence between real distribution ${{p}({ \boldsymbol x_{age} |  \boldsymbol x_{id}})}$ and the variational distribution $q_{\theta}({ \boldsymbol x_{age}}|{ \boldsymbol x_{id}})$ so that $q_{\theta}({ \boldsymbol x_{age}}|{ \boldsymbol x_{id}})$ is an accurate approximation to ${{p}({ \boldsymbol x_{age} |  \boldsymbol x_{id}})}$:
\begin{equation}
    \label{eq7}
\mathcal{L}_{KL}=KL(p( \boldsymbol x_{age}| \boldsymbol x_{id})\parallel q_{\theta}( \boldsymbol x_{age}| \boldsymbol x_{id}))
\end{equation}
Since the absolute error is bounded by the approximation performance $KL(p( \boldsymbol x_{age}| \boldsymbol x_{id})\parallel q_{\theta}( \boldsymbol x_{age}| \boldsymbol x_{id}))$, the MI between  $ \boldsymbol x_{id}$ and  $ \boldsymbol x_{age}$ is also estimated with less bias by maximizing the $\mathbb{E}_{p(\mathbf{x}_{id},\mathbf{x}_{agc})}[\log q_{\theta}(\mathbf{x}_{age}|\mathbf{x}_{id})]$ .

\subsubsection{Aging-Aware Mutual Information Minimization}
In CASV, since age labels are noisy and aging factors have different effects on people, $\mathcal{L}_{MIM}$ fluctuates within a very wide range due to
the log-ratio bound. We propose to replace the log-ratio bound with probability-ratio so that $\mathcal{L}_{MIM}$ fluctuates more slightly.

Moreover, in a previous study~\cite{reubold2010vocal}, the researchers found that vocal aging across short-term is not obvious in measurable acoustic parameters~(i.e., F0). This is also consistent with human perception, where vocal changes over a few years are almost imperceptible. In Eq.~\ref{eq4}, the negative sample $\boldsymbol{x}_{age}^{k_{i}^{\prime}}$ is randomly selected from the same batch. However, $q_{\theta}(\boldsymbol{x}_{age}^{k_{i}^{\prime}}|\boldsymbol{x}_{id}^{i})$ is expected to be equally maximized regardless of the age gaps between $\boldsymbol{x}_{age}^{k_{i}^{\prime}}$ and $\boldsymbol{x}_{age}^{i}$. 

To better utilize the relations contained in age labels, we reweight the age gap of $\boldsymbol{x}_{age}^{i}$ and $\boldsymbol{x}_{age}^{k_{i}^{\prime}}$ to learn correlations among different ages by a hyper-parameter $ \lambda_i^{{k_{i}^{\prime}}}$. This aging-aware objective function is defined as:
\begin{equation}
    \label{eq5}
\lambda_i^{{k_{i}^{\prime}}} = \log (|z_{age}^i - z_{age}^{{k_{i}^{\prime}}}|+\lambda_0)
\end{equation}
\begin{equation}
    \label{eq6}
\mathcal{L}_{AA-MIM}=\frac{1}{N}\sum_{i=1}^{N}\Big[ q_{\theta}(\boldsymbol{x}_{age}^{i}|\boldsymbol{x}_{id}^{i})- \lambda_i^{{k_{i}^{\prime}}}q_{\theta}(\boldsymbol{x}_{age}^{k_{i}^{\prime}}|\boldsymbol{x}_{id}^{i})\Big]
\end{equation}
where $z_{age}^i, z_{age}^{{k_{i}^{\prime}}}$ is the ground truth age labels and $ \lambda_0 $ is an offset hyper-parameter. Via Eq.~\ref{eq6}, the backbone model is trained to minimize the MI between $ \boldsymbol x_{id}$ and $ \boldsymbol x_{age}$.

\subsection{Overall Framework}
As illustrated in Fig.~\ref{fig2}, the backbone model with parameter $ \phi$ is optimized by three supervised tasks: identity classification loss $\mathcal{L}_{id}$, age classification loss $ \mathcal{L}_{age}$ , MI minimization loss $\mathcal{L}_{AA-MIM}$:
\begin{equation}
    \label{eq10}
\mathcal{L}_{\phi}=\mathcal{L}_{id}+\lambda_{age}\mathcal{L}_{age}+\lambda_{MI}\mathcal{L}_{AA-MIM}
\end{equation}
where $\lambda_{age}$ and $ \lambda_{MI}$ are hyper-parameters to balance the overall loss. 
The overall training process is described in Algorithm~\ref{alg:algorithm1}.
\begin{algorithm}[h]
	\caption{Overall Training Process}
	\label{alg:algorithm1}
	\KwIn{Training data $\{x^i\}_{i=1}^N$, \newline The backbone model with parameters $\phi$, \newline MI estimator with parameters $\theta$, \newline Epoch numbers $ E$, \newline Batch numbers $ N$.}
	\KwOut{Trained $\phi$ and $\theta$.}  
	\BlankLine
	Initialize $\phi$ and $\theta$ randomly;
 
    \ForEach{ epoch }
    {
    {\ForEach{ batch }
    { 
    $\phi$ optimization~($\theta$ fixed):
    \begin{itemize}
\item[]
        Obtain $ \boldsymbol x_{age}^i,  \boldsymbol x_{id}^i$ via forward propagation $ \phi(x^i)$ ;
\end{itemize}
\begin{itemize}
\item[]
        Compute $\mathcal{L}_{\mathrm{id}}$ using $ \boldsymbol x_{id}^i$;
\end{itemize}
\begin{itemize}
\item[]
        Compute $\mathcal{L}_{\mathrm{age}}$ using $ \boldsymbol x_{age}^i$;
\end{itemize}
\begin{itemize}
\item[]
        Compute $\mathcal{L}_{\mathrm{AA-MIM}}$ using $ \boldsymbol x_{age}^i$ and $ \boldsymbol x_{id}^i$ ;
\end{itemize}
\begin{itemize}
\item[]
        Compute $\mathcal{L}_{\phi}$ by Eq.~\ref{eq10} ;
\end{itemize}
\begin{itemize}
\item[]
Update $\phi$ via back-propagation.
\end{itemize}  
    
    { $\theta$ optimization~($\phi$ fixed):
    \begin{itemize}
\item[]
        Compute $\mathcal{L}_{\theta}$ by Eq.~\ref{eq7} ;
\end{itemize}
\begin{itemize}
\item[]
Update $\theta$ via back-propagation.
\end{itemize} 
    }}
    }
    }
    \Return $\phi$ and $\theta$.
\end{algorithm}

\section{Experiments}
\subsection{Datasets}
Vox-CA train set ~\cite{qin22_interspeech} constructed on Voxceleb2~\cite{chung18b_interspeech} is used to train the model which contains 5990 speakers with 1085425 utterances. Vox-CA test sets constructed on Voxceleb1~\cite{nagrani17_interspeech} consist of multiple test trials with different cross-age scenes. We chose both Only-CA and Vox-CA trials for evaluation. Compared with Only-CA, Vox-CA takes gender and nationality into account when constructing negative pairs. In addition, the widely used Voxceleb1 test sets(i.e., Vox-O, Vox-E, Vox-H) are adopted to evaluate the performance of the general speaker verification task. The details of the test sets used in this paper are reported in Table.~\ref{dataset}. MUSAN~\cite{snyder2015musan} and RIRs~\cite{ko2017study} are used for data augmentation.
\begin{table}[t!]
\begin{center}
\caption{Details of the test sets. Part from~\cite{qin22_interspeech}. Spk.Num and Trials Num denote the number of enrolled speakers and trial pairs, respectively. Positive and Negative denote the mean and standard deviation of age gap corresponding to trial pairs.}
\resizebox{\linewidth}{!}{
\begin{tabular}{lcccc}
\hline
\label{dataset}
 \multirow{2}{*}{Test set} &  \multirow{2}{*}{Spk.Num}& \multirow{2}{*}{Trials Num}& \multicolumn{2}{c}{Age-gap}  \\ 
 \cmidrule(r){4-5}
 &&&Positive& Negative\\\hline
Vox-O& 40 & 37611 & 2.68  $\pm$ 2.88 & 15.50$\pm$12.46\\
Vox-E& 1251 & 579818 & 3.14$\pm$3.48 & 12.05$\pm$9.81\\
Vox-H& 1190& 550894 & 3.14$\pm$3.47&11.27$\pm$9.42 \\ \hline
Only-CA5& 1013 & 486404 & 7.98$\pm$3.378 & 12.40$\pm$9.83\\
Only-CA10& 533 & 158332 & 15.29$\pm$3.46 &15.43$\pm$10.30\\
Only-CA15& 224 & 56792 & 20.43$\pm$ 3.37 & 17.65$\pm$10.62\\
Only-CA20& 89 & 19940 & 25.25$\pm$2.81 &19.30$\pm$10.84 \\ \hline
Vox-CA5& 971 & 370540 & 9.98$\pm$3.94 & 12.36$\pm$9.58\\
Vox-CA10& 506 & 151384 & 15.29$\pm$3.44 & 14.66$\pm$9.93\\
Vox-CA15& 215 & 54608 & 20.39$\pm$ 3.38 & 16.63$\pm$10.24\\
Vox-CA20& 85 & 18888 & 25.28$\pm$2.87 & 18.42$\pm$10.58 \\
\hline
\end{tabular}
}
\end{center}
\vspace{-0.4cm}
\end{table}
\begin{table*}[t]
\begin{center}
\caption{EER~(\%) / minDCF on Vox-CA test sets.* denotes the official results reported in~\cite{qin22_interspeech}. }
\label{result}
\resizebox{\linewidth}{!}{
\begin{tabular}{lcccccccc | c} 
\hline
 \multirow{2}{*}{Method} &  \multicolumn{4}{c}{Cross-age} &  \multicolumn{4}{c|}{Cross-age \& Same nationality \& Same gender} & \multirow{2}{*}{Average} \\ 
 \cmidrule(r){2-5} \cmidrule(r){6-9}
 &Only-CA5&Only-CA10&Only-CA15&Only-CA20&Vox-CA5&Vox-CA10&Vox-CA15& Vox-CA20&\\
 \hline 
ResNet34~\cite{he2016deep}& 1.958 / 0.173& 3.28 / 0.271&5.596 / 0.375 & 7.803 / 0.458 & 3.575 / 0.294&  5.179 / 0.360 & 8.167 / 0.489 & 10.631 / 0.658 & 5.774 / 0.385\\
ResNet34+ADAL~\cite{qin22_interspeech}& 1.901 / \textbf{0.163}& 3.139 / \textbf{0.253}& \textbf{5.149} / 0.384&\textbf{7.442} / 0.454 & 3.473 / 0.285 & 4.907 / 0.364& 7.867 / 0.484& 10.695 / \textbf{0.622}& 5.572 / 0.376 \\
\textbf{Ours w/o MIM}& 2.041 / 0.180& 3.376 / 0.283 & 5.434 / 0.359  & 7.603 / 0.470 & 3.657 / 0.302 & 5.314 / 0.375 & 8.398 / 0.509 & 10.864 / 0.658 & 5.836 / 0.392 \\

\textbf{Ours w/o AA} & 1.887 / 0.174& 3.123 / 0.248 & 5.473 / \textbf{0.319}& 7.503 / \textbf{0.448} & 3.417 / 0.272 & 4.831 / 0.339 & 7.819 / 0.430& 10.388 / 0.644 & 5.556 / 0.359 \\
\textbf{Ours}& \textbf{1.857} / 0.184& \textbf{3.076} / 0.249 &5.406 / 0.320 &7.492 / 0.451 & \textbf{3.386} / \textbf{0.271} & \textbf{4.801} / \textbf{0.336}&\textbf{7.72} / \textbf{0.424} & \textbf{10.155} / 0.626&  \textbf{5.487} / \textbf{0.358}\\
\hline
*ResNet34~\cite{he2016deep} & 1.953 / 0.177& 3.437 / 0.272 & 5.927 / 0.352 & 8.185 / 0.464 & 3.407 / 0.300 & 4.974 / 0.370 & 8.028 / 0.481 & 10.419 / 0.646 & 5.791 /  0.383\\
*ResNet34+ADAL~\cite{qin22_interspeech} & 1.991 / -&3.33 / - &5.54 / - & 7.442 / -& 3.441 / - & 4.822 / -&7.515 / - & 9.519 / - & 5.45 / -\\
\hline 
\label{results}
\end{tabular} 
}
\end{center}
  \vspace{-0.6cm}
\end{table*}
\begin{table}[t]
\begin{center}
\caption{EER~(\%) / minDCF on Voxceleb official test sets.}
\resizebox{\linewidth}{!}{
\begin{tabular}{lc cc}
\hline
\label{general}
Method&Vox-O&Vox-E& Vox-H\\
\hline
ResNet34& \textbf{0.899} / 0.099 & 1.064 / 0.122 & 1.971 / 0.194 \\
ResNet34+ADAL& 0.925 / 0.106 & 1.030 / 0.116 & 1.873 / \textbf{0.181} \\
Ours& 0.936 / \textbf{0.066} & \textbf{1.008} / \textbf{0.114} & \textbf{1.858} / 0.184 \\
\hline
\end{tabular}}
\end{center}

\end{table}
\subsection{Experimental Settings}
The input 80-dimensional Fbank is extracted with a frame length of 25ms and frame hop of 10ms. Three data augmentation methods are adopted: 1) adding noise using MUSAN. 2) adding reverberation using RIR. 3) speed perturbation. When training, each utterance is cut into chunks that contain a length of 200 frames.

For the backbone model, Resnet34~\cite{he2016deep} is adopted as the feature encoder. The initial learning rate~(LR) of the Stochastic gradient descent (SGD) optimizer is 0.1 and follows an ExponentialDecrease with a weight decay of 1e-4. We also adopt the warming up LR in the first six epochs. The training process is stopped when LR dropped to 5e-5. The ArcFace loss $\mathcal{L}_{id}$ is used with a scaling factor of 48 and a margin of 0.2. The softmax loss $\mathcal{L}_{age}$ is used for age group classification, where age is split into 7 groups: 0-20, 21-30, 31-40, 41-50, 51-60, 61-70, 71-80.
The hyper-parameters of $\mathcal{L}_{\phi}$ are set as following: $\lambda_{age} = 0.1$ and $\lambda_{MI} = 0.0001$. 

For the MI estimator, $ {q_\theta}({ \boldsymbol x_{age}|{{ \boldsymbol x_{id}}}})$ is parameterized by a Gaussian distribution. The mean and variance vectors are obtained by a series of fully-connected (FC) layers and non-linear transformations. The network to obtain the mean vector consists of a $256 \times 512$ FC layer, a ReLU layer and a $512 \times 256$ FC layer. The network to obtain the variance vector consists of a $256 \times 512$ FC layers, a ReLU layer, a $512 \times 256$ FC layer and a Tanh layer. Adam~\cite{kingma2014adam} with an LR of 1e-5 and a weight decay of 1e-4 is used to optimize the MI estimator.

For all networks, the batch size of the input is 384. No back-end scoring is used in inference. The experiment is conducted on Wespeaker~\cite{wang2023wespeaker}.

Same as most SV tasks, Equal Error Rate (EER) and Minimum Detection Cost Function (minDCF) with $P_{target}=0.01$ and $C_{FA}=C_{Miss}=1$ are adopted as the evaluation metrics.

\section{Results}
\subsection{Results on Cross-age tasks}

Table.~\ref{result} shows the performance of the current methods on Vox-CA test sets. We re-implement ResNet34~\cite{he2016deep} and SOTA ResNet34+ADAL~\cite{qin22_interspeech} for comparison. 
Compared with ResNet34+ADAL, our method achieves relative improvement in overall performance by 1.53\% in EER and 4.79\% in minDCF, respectively, demonstrating that our method learns better age-invariant speaker embeddings. 

In addition, we further conduct two ablation experiments to analyze the contributions of each proposed component. `Ours w/o AA' denotes that the aging-aware function is removed and $ \lambda_i^{{k_{i}^{\prime}}}$ is set to 1 regardless of the age gaps in Eq.\ref{eq5}. Then if the MI estimator is removed~(`Ours w/o MIM'), the backbone model is only supervised by age and identity classification loss. As can be seen in Table.~\ref{result}, without the aging-aware function, the performance declines on most sets. Then if the MIM process is removed, the performance becomes even worse than the baseline ResNet34 model. 
We consider that the architecture of the backbone model to split $ \boldsymbol x_{id}$ and $ \boldsymbol x_{age}$ from $ \boldsymbol x_{init}$ is effective only when combined with disentanglement methods~(i.e., gradient reversal~\cite{ganin2015unsupervised} and MI minimization~\cite{cheng2020club}). Without disentanglement methods to model the relations between age and identity, the performance can not be improved.

\subsection{Results on General tasks}
In this subsection, we introduce the performance of Voxceleb official test sets, which are conducted for general speaker verification. As is reported in Table.~\ref{general}, the results among the three methods are similar, which indicates that the disentanglement of age- and identity-related information has little influence on general speaker verification.


\begin{figure}[t]
  \centering
  \includegraphics[width=\linewidth]{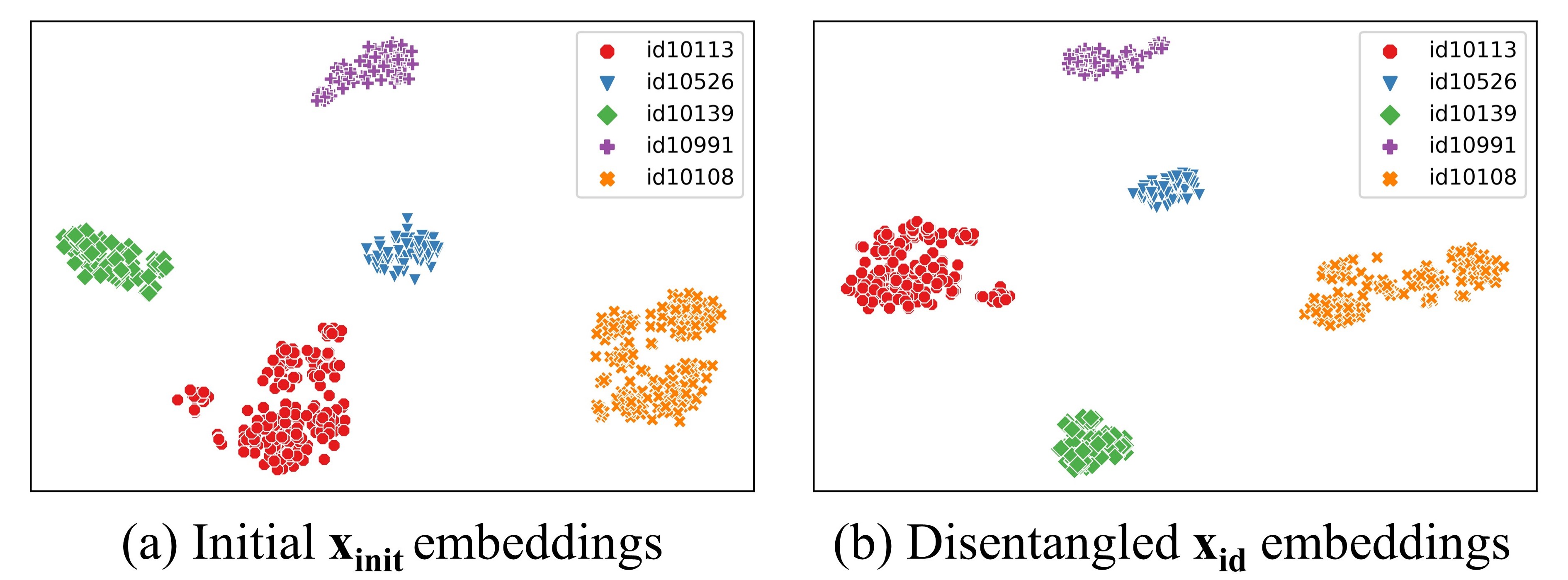}
  \caption{t-SNE plots of initial and disentangled embeddings.
  }
  \label{fig3}

\end{figure}

\subsection{Visualization of Disentanglement}
To qualitatively illustrate the effectiveness of the disentanglement, we showcase t-SNE of speaker embeddings in the cross-age scenes. We chose the target speaker 'id10113' and its negative trials in Vox-CA20 for visualization. As can be seen in Fig.~\ref{fig3}, the points with red color represent the speaker embedding from the target speaker, while others denote the negative speakers. Since the initial $ \boldsymbol x_{init}$ contains both age and identity information, resulting in great intra-identity distance, the target speaker can be easily rejected. Compared with $ \boldsymbol x_{init}$, the disentangled $ \boldsymbol x_{id}$ embeddings show lower intra-identity distance, which indicates that $ \boldsymbol x_{id}$ tends to be age-invariant.
\section{Conclusion}

In this paper, we propose a disentangling method for CASV based on MIM. The innovation of our proposed method lies in introducing an MI estimator to the backbone model to disentangle age and identity information.
By minimizing the MI between age and identity embedding, our method leads to better age-invariant speaker embeddings. 
In addition, we proposed an aging-aware MIM loss to utilize the relations between different ages. Experimental results in Vox-CA test sets outperform the SOTA method by 1.53\% in EER and 4.79\% in minDCF, respectively, demonstrating the effectiveness of our method.

\bibliographystyle{IEEEtran}
\bibliography{mybib}

\begin{thebibliography}{10}
\providecommand{\url}[1]{#1}
\csname url@samestyle\endcsname
\providecommand{\newblock}{\relax}
\providecommand{\bibinfo}[2]{#2}
\providecommand{\BIBentrySTDinterwordspacing}{\spaceskip=0pt\relax}
\providecommand{\BIBentryALTinterwordstretchfactor}{4}
\providecommand{\BIBentryALTinterwordspacing}{\spaceskip=\fontdimen2\font plus
\BIBentryALTinterwordstretchfactor\fontdimen3\font minus \fontdimen4\font\relax}
\providecommand{\BIBforeignlanguage}[2]{{%
\expandafter\ifx\csname l@#1\endcsname\relax
\typeout{** WARNING: IEEEtran.bst: No hyphenation pattern has been}%
\typeout{** loaded for the language `#1'. Using the pattern for}%
\typeout{** the default language instead.}%
\else
\language=\csname l@#1\endcsname
\fi
#2}}
\providecommand{\BIBdecl}{\relax}
\BIBdecl

\bibitem{snyder2018x}
D.~Snyder, D.~Garcia-Romero, G.~Sell, D.~Povey, and S.~Khudanpur, ``X-vectors: Robust dnn embeddings for speaker recognition,'' in \emph{2018 IEEE international conference on acoustics, speech and signal processing (ICASSP)}.\hskip 1em plus 0.5em minus 0.4em\relax IEEE, 2018, pp. 5329--5333.

\bibitem{desplanques20_interspeech}
B.~Desplanques, J.~Thienpondt, and K.~Demuynck, ``{ECAPA-TDNN: Emphasized Channel Attention, Propagation and Aggregation in TDNN Based Speaker Verification},'' in \emph{Proc. Interspeech 2020}, 2020, pp. 3830--3834.

\bibitem{chen23o_interspeech}
Y.~Chen, S.~Zheng, H.~Wang, L.~Cheng, Q.~Chen, and J.~Qi, ``{An Enhanced Res2Net with Local and Global Feature Fusion for Speaker Verification},'' in \emph{Proc. Interspeech 2023}, 2023, pp. 2228--2232.

\bibitem{deng2019arcface}
J.~Deng, J.~Guo, N.~Xue, and S.~Zafeiriou, ``Arcface: Additive angular margin loss for deep face recognition,'' in \emph{Proceedings of the IEEE/CVF conference on computer vision and pattern recognition}, 2019, pp. 4690--4699.

\bibitem{zhao2022multi}
M.~Zhao, Y.~Ma, Y.~Ding, Y.~Zheng, M.~Liu, and M.~Xu, ``Multi-query multi-head attention pooling and inter-topk penalty for speaker verification,'' in \emph{ICASSP 2022-2022 IEEE International Conference on Acoustics, Speech and Signal Processing (ICASSP)}.\hskip 1em plus 0.5em minus 0.4em\relax IEEE, 2022, pp. 6737--6741.

\bibitem{qin20_interspeech}
X.~Qin, M.~Li, H.~Bu, W.~Rao, R.~K. Das, S.~Narayanan, and H.~Li, ``{The Interspeech 2020 Far-Field Speaker Verification Challenge},'' in \emph{Proc. Interspeech 2020}, 2020, pp. 3456--3460.

\bibitem{thienpondt2022tackling}
J.~Thienpondt, B.~Desplanques, and K.~Demuynck, ``Tackling the score shift in cross-lingual speaker verification by exploiting language information,'' in \emph{ICASSP 2022-2022 IEEE International Conference on Acoustics, Speech and Signal Processing (ICASSP)}.\hskip 1em plus 0.5em minus 0.4em\relax IEEE, 2022, pp. 7187--7191.

\bibitem{gonzalez2019limits}
R.~Gonz{\'a}lez~Hautam{\"a}ki, V.~Hautam{\"a}ki, and T.~Kinnunen, ``On the limits of automatic speaker verification: Explaining degraded recognizer scores through acoustic changes resulting from voice disguise,'' \emph{The Journal of the Acoustical Society of America}, vol. 146, no.~1, pp. 693--704, 2019.

\bibitem{tawara2021age}
N.~Tawara, A.~Ogawa, Y.~Kitagishi, and H.~Kamiyama, ``Age-vox-celeb: Multi-modal corpus for facial and speech estimation,'' in \emph{ICASSP 2021-2021 IEEE International Conference on Acoustics, Speech and Signal Processing (ICASSP)}.\hskip 1em plus 0.5em minus 0.4em\relax IEEE, 2021, pp. 6963--6967.

\bibitem{hechmi2021voxceleb}
K.~Hechmi, T.~N. Trong, V.~Hautam{\"a}ki, and T.~Kinnunen, ``Voxceleb enrichment for age and gender recognition,'' in \emph{2021 IEEE Automatic Speech Recognition and Understanding Workshop (ASRU)}.\hskip 1em plus 0.5em minus 0.4em\relax IEEE, 2021, pp. 687--693.

\bibitem{nagrani17_interspeech}
A.~Nagrani, J.~S. Chung, and A.~Zisserman, ``{VoxCeleb: A Large-Scale Speaker Identification Dataset},'' in \emph{Proc. Interspeech 2017}, 2017, pp. 2616--2620.

\bibitem{chung18b_interspeech}
J.~S. Chung, A.~Nagrani, and A.~Zisserman, ``{VoxCeleb2: Deep Speaker Recognition},'' in \emph{Proc. Interspeech 2018}, 2018, pp. 1086--1090.

\bibitem{qin22_interspeech}
X.~Qin, N.~Li, W.~Chao, D.~Su, and M.~Li, ``{Cross-Age Speaker Verification: Learning Age-Invariant Speaker Embeddings},'' in \emph{Proc. Interspeech 2022}, 2022, pp. 1436--1440.

\bibitem{singh23d_interspeech}
V.~P. Singh, M.~Sahidullah, and T.~Kinnunen, ``{Speaker Verification Across Ages: Investigating Deep Speaker Embedding Sensitivity to Age Mismatch in Enrollment and Test Speech},'' in \emph{Proc. Interspeech 2023}, 2023, pp. 1948--1952.

\bibitem{mueller1997aging}
P.~B. Mueller, ``The aging voice,'' in \emph{Seminars in speech and language}, vol.~18, no.~02.\hskip 1em plus 0.5em minus 0.4em\relax {\copyright} 1997 by Thieme Medical Publishers, Inc., 1997, pp. 159--169.

\bibitem{reubold2010vocal}
U.~Reubold, J.~Harrington, and F.~Kleber, ``Vocal aging effects on f0 and the first formant: A longitudinal analysis in adult speakers,'' \emph{Speech communication}, vol.~52, no. 7-8, pp. 638--651, 2010.

\bibitem{goodfellow2014generative}
I.~Goodfellow, J.~Pouget-Abadie, M.~Mirza, B.~Xu, D.~Warde-Farley, S.~Ozair, A.~Courville, and Y.~Bengio, ``Generative adversarial nets,'' \emph{Advances in neural information processing systems}, vol.~27, 2014.

\bibitem{hou2021disentangled}
X.~Hou, Y.~Li, and S.~Wang, ``Disentangled representation for age-invariant face recognition: A mutual information minimization perspective,'' in \emph{Proceedings of the IEEE/CVF International Conference on Computer Vision}, 2021, pp. 3692--3701.

\bibitem{wang2021learning}
Z.~Wang, Y.~Luo, R.~Qiu, Z.~Huang, and M.~Baktashmotlagh, ``Learning to diversify for single domain generalization,'' in \emph{Proceedings of the IEEE/CVF International Conference on Computer Vision}, 2021, pp. 834--843.

\bibitem{li23r_interspeech}
J.~Li, J.~Han, S.~Deng, T.~Zheng, Y.~He, and G.~Zheng, ``{Mutual Information-based Embedding Decoupling for Generalizable Speaker Verification},'' in \emph{Proc. INTERSPEECH 2023}, 2023, pp. 3147--3151.

\bibitem{cheng2020club}
P.~Cheng, W.~Hao, S.~Dai, J.~Liu, Z.~Gan, and L.~Carin, ``Club: A contrastive log-ratio upper bound of mutual information,'' in \emph{International conference on machine learning}.\hskip 1em plus 0.5em minus 0.4em\relax PMLR, 2020, pp. 1779--1788.

\bibitem{snyder2015musan}
D.~Snyder, G.~Chen, and D.~Povey, ``Musan: A music, speech, and noise corpus,'' \emph{arXiv preprint arXiv:1510.08484}, 2015.

\bibitem{ko2017study}
T.~Ko, V.~Peddinti, D.~Povey, M.~L. Seltzer, and S.~Khudanpur, ``A study on data augmentation of reverberant speech for robust speech recognition,'' in \emph{2017 IEEE International Conference on Acoustics, Speech and Signal Processing (ICASSP)}.\hskip 1em plus 0.5em minus 0.4em\relax IEEE, 2017, pp. 5220--5224.

\bibitem{he2016deep}
K.~He, X.~Zhang, S.~Ren, and J.~Sun, ``Deep residual learning for image recognition,'' in \emph{Proceedings of the IEEE conference on computer vision and pattern recognition}, 2016, pp. 770--778.

\bibitem{kingma2014adam}
D.~P. Kingma and J.~Ba, ``Adam: A method for stochastic optimization,'' \emph{arXiv preprint arXiv:1412.6980}, 2014.

\bibitem{wang2023wespeaker}
H.~Wang, C.~Liang, S.~Wang, Z.~Chen, B.~Zhang, X.~Xiang, Y.~Deng, and Y.~Qian, ``Wespeaker: A research and production oriented speaker embedding learning toolkit,'' in \emph{ICASSP 2023-2023 IEEE International Conference on Acoustics, Speech and Signal Processing (ICASSP)}.\hskip 1em plus 0.5em minus 0.4em\relax IEEE, 2023, pp. 1--5.

\bibitem{ganin2015unsupervised}
Y.~Ganin and V.~Lempitsky, ``Unsupervised domain adaptation by backpropagation,'' in \emph{International conference on machine learning}.\hskip 1em plus 0.5em minus 0.4em\relax PMLR, 2015, pp. 1180--1189.

\end{thebibliography}

\end{document}